\documentclass[a4paper,onecolumn,english,11pt,amssymb,nofootinbib,superscriptaddress]{revtex4-2}
\usepackage{lmodern}
\usepackage{lmodern}

\usepackage[T1]{fontenc}
\usepackage[latin9]{inputenc}
\usepackage{babel}
\usepackage{mathrsfs}
\usepackage{amsmath}
\usepackage{amssymb}
\usepackage{esint}
\usepackage{dsfont}
\usepackage[unicode=true,pdfusetitle,
bookmarks=true,bookmarksnumbered=false,bookmarksopen=false,
breaklinks=false,pdfborder={0 0 1},backref=false,colorlinks=false]
{hyperref}

\makeatletter

\@ifundefined{textcolor}{}
{%
	\definecolor{BLACK}{gray}{0}
	\definecolor{WHITE}{gray}{1}
	\definecolor{RED}{rgb}{1,0,0}
	\definecolor{GREEN}{rgb}{0,1,0}
	\definecolor{BLUE}{rgb}{0,0,1}
	\definecolor{CYAN}{cmyk}{1,0,0,0}
	\definecolor{MAGENTA}{cmyk}{0,1,0,0}
	\definecolor{YELLOW}{cmyk}{0,0,1,0}
}

\usepackage{babel}
\usepackage{babel}
\usepackage{babel}
\usepackage{babel}
\usepackage{babel}
\usepackage{babel}
\usepackage{babel}
\usepackage{babel}
\usepackage{babel}
\usepackage{babel}
\usepackage{babel}
\usepackage{babel}
\usepackage{babel}
\usepackage{babel}
\usepackage{babel}
\usepackage{babel}
\usepackage{babel}
\usepackage{graphicx}
\def\b{\begin{equation}}
\def\e{\end{equation}}

\@ifundefined{textcolor}{}{%
	\definecolor{BLACK}{gray}{0}
	\definecolor{WHITE}{gray}{1}
	\definecolor{RED}{rgb}{1,0,0}
	\definecolor{GREEN}{rgb}{0,1,0}
	\definecolor{BLUE}{rgb}{0,0,1}
	\definecolor{CYAN}{cmyk}{1,0,0,0}
	\definecolor{MAGENTA}{cmyk}{0,1,0,0}
	\definecolor{YELLOW}{cmyk}{0,0,1,0}
}

\usepackage{latexsym}\usepackage{bm}

\makeatother

\begin{document}
	\title{Partition function of a volume of space in a higher curvature theory}
	
	\author{Aydin Tavlayan}
	
	\email{aydint@metu.edu.tr}
	
	\selectlanguage{english}%
	
	\affiliation{Department of Physics,\\
		Middle East Technical University, 06800 Ankara, Turkey}
	\author{Bayram Tekin}
	
	\email{btekin@metu.edu.tr}
	\affiliation{Department of Physics,\\
		Middle East Technical University, 06800 Ankara, Turkey}

	\selectlanguage{english}%
\begin{abstract}
Recently, [Phys. Rev. Lett. \textbf{130}, 221501 (2023)] Jacobson and Visser calculated the quantum partition function of a fixed, finite volume of a region with the topology of a ball in the saddle point approximation within the context of Einstein's gravity with or without a cosmological constant.  The result can be interpreted as the dimension of Hilbert space of the theory. Here we extend their computation to a theory defined in principle with infinitely many powers of curvature in three dimensions. We confirm their result: The partition function of a spatial region in the leading saddle point approximation is given as the exponential of the Bekenstein-Hawking or the Wald entropy of the boundary of the finite spatial region both in the case of zero and finite cosmological constant.  In the latter case, the effective Newton's constant appears in the entropy formula. The calculations lend support to the holographic nature of gravity for finite regions of space with a boundary.
\end{abstract}
\maketitle

\section{Introduction}
Over the last five decades, we have come to understand that the gravitational field has typically a finite (nonzero) entropy;  but which kinds of gravitational fields have entropy and how this entropy is defined is a subtle problem. However, the fact that entropy has both macroscopic and microscopic definitions, and its microscopic description is most probably tied to the quantum version of the theory, makes it an invaluable tool in understanding some properties of low energy quantum gravity. In the original incarnation \cite{Bekenstein1, Hawking1}, the entropy of the gravitational field was assigned to {\it black hole horizons}. For example, the Bekenstein-Hawking entropy of  
a Schwarzschild black hole is
\begin{equation}
S_{BH} = \frac{ k_B c^3}{4 G\hbar} A_H,
\end{equation}
where $A_H$ is the area of the event horizon.
Gibbons and Hawking \cite{GH} extended this result to the case of {\it de Sitter horizons}
\begin{equation}
S_{GH} = \frac{ k_B c^3}{4 G\hbar} A_{dS}= \frac{3\pi k_B c^3}{G\hbar \Lambda} ,
\end{equation}
where $A_{dS}$ is the area of the cosmological horizon located at $r_H= \sqrt{3/ \Lambda}$ in the usual coordinates. Note also that $\Lambda$ is defined as $\Lambda = R/4$  with $R$ being the constant scalar curvature of the de Sitter spacetime.  In the same work, Gibbons and Hawking argue that the gravitational field of a stationary star has zero entropy.  
The physical meaning of the de Sitter entropy is not so clear; for example, in \cite{Bala} seven possible interpretations were given.  Here we will follow the interpretation advocated in \cite{Fischler,Banks} which states that de Sitter entropy is equal to the logarithm of the dimension of the Hilbert space. One can understand the viability of this interpretation as follows: The trace of the unit operator ($\text{Tr}\, \mathds{1}$) which, for a finite-dimensional Hilbert space,  is equal to the dimension of the Hilbert space. For de Sitter space, which does not have a boundary, the total Hamiltonian vanishes identically and one has ($\text{dim}{\mathcal{H}}=\text{Tr}\, \mathds{1} = \text{Tr} e^{- \beta \hat{H} }$). The right-hand side at the same time is the partition function, which can be evaluated in the Euclidean path integral formulation. The new and exciting development in this subject is the work of Jacobson and Visser \cite{Visser1} where the entropy of a spatial region of space with a fixed proper volume and a boundary was defined in Einstein's theory with or without a cosmological constant. See \cite{Visser2} for more details.  They show that in the saddle point approximation, the quantum gravity path integral under the fixed volume condition is dominated by the so-called constrained instantons \cite{Affleck, Cotler} of which the Euclidean action is minus the Bekenstein-Hawking entropy calculated for the area bounding the spatial volume.  As long as Einstein's gravity is correct as an effective field theory, this computation is robust with only one caveat: There is a mild singularity in the Ricci tensor and the energy-momentum tensor in the boundary which does not make the action divergent. This divergence is expected to be cured when higher powers of curvature are added. This point motivated us to carry out an analogous computation in a higher curvature theory which has rather nice properties and extends Einstein's gravity to infinitely many curvatures. The mild singularity observed in \cite{Visser1} is not cured, but the results of the computations lend strong support to   \cite{Visser1}.

\section{Partition Function for a Fixed Volume of Space}
The quantum gravity partition function with a constraint in the general form can be written as
\begin{eqnarray}
\mathcal{Z}&=&\int {\cal{D}}g e^{-I_E(g)} = \int  {\cal{D}}\mu \int {\cal{D}}\xi \int  {\cal{D}}g e^{-I_{E}(g)+\frac{1}{\hbar}\int d\phi \xi(\phi)\left(C(g)-\mu\right)},
\end{eqnarray}
where $I_{E}(g)$ represents the Euclidean action that is under investigation, $\xi$ is the Lagrange multiplier, and $C(g)-\mu=0$ is the constraint equation. We shall consider the three-dimensional gravity introduced in \cite{Tekin1}. In $2+1$ dimensions, Einstein's gravity with or without a cosmological constant has no local degrees of freedom. Hence, it has no resemblance to four-dimensional Einstein theory. In \cite{NMG}, a massive three-dimensional gravity theory was introduced which has two propagating degrees of freedom, just like four-dimensional Einstein gravity. The action of this new massive gravity theory (NMG) is 
\begin{eqnarray}
I_{NMG} &=&  \frac{1}{\kappa^2} \int d^3 x \sqrt{-\det\left(g\right)}\left[\sigma R +\frac{1}{m^2}\left(R_{\mu\nu}R^{\mu\nu}-\frac{3}{8}R^2\right)-2\lambda_0 m^2\right],
\end{eqnarray}
where $\sigma = \pm 1$.  Here, $\lambda_0$ is the bare dimensionless cosmological constant.
In \cite{Tekin1}, this theory was extended in principle to infinite powers in the curvature expansion, and the new action is defined as
\begin{eqnarray}
I_{BI-NMG} &=& - \frac{4 m^2}{\kappa^2} \int d^3 x \left[\sqrt{-\det\left(g+\frac{\sigma}{m^2}G\right)}- \left(1-\frac{\lambda_0}{2}\right) \sqrt{-\det\left(g\right)} \right], \label{BINMG}
\end{eqnarray}
where $G_{\mu\nu} = R_{\mu\nu}-\frac{1}{2}g_{\mu\nu}R$. This theory has the following rather nice properties which we take from \cite{Deser-Tekin}.
\begin{enumerate}
\item For $\lambda_0 = 0$, unlike any generic finite-order theory besides Einstein's gravity with a cosmological constant, it has a {\it unique} maximally symmetric vacuum with an effective cosmological constant $\Lambda=m^2\lambda$ given as \cite{Tekin2,Nam}
\begin{equation}
\lambda = -\sigma \lambda_0 \left(1-\frac{\lambda_0}{4}\right), \,\,\,\lambda_0<2.
\end{equation}
For $\lambda_0 = 0$, the flat space is the unique vacuum. The uniqueness of the vacuum in this higher derivative theory is the same as in general relativity. This uniqueness property is not easily achievable in higher curvature gravity theories: For example, NMG has two maximally symmetric vacua.
\item The theory has a unitary massive spin-2 degree of freedom (with $\pm 2$ modes) in flat space with mass $m_g=m$ and with mass $m_g = m \sqrt{1 + \lambda}$ around the anti-de Sitter (AdS) backgrounds. This provides an infinite-order extension of the quadratic NMG which has the same perturbative properties.
\item In three dimensions, the Riemann and the Ricci tensors both carry six independent components; hence, a generic gravity theory built from the contractions and the powers of the Ricci curvature is of the form \cite{Gurses}
\begin{equation}
I=\int d^3 x \sqrt{-g}\left[\frac{1}{\kappa}\left(R-2 \Lambda_0\right)+\sum_{n=2}^{\infty} \sum_{\substack{i=0 \\ i \neq 1}}^n \sum_j a_{n i}^j\left(R_\nu^\mu\right)_j^i R^{n-i}\right], \label{Gen}
\end{equation}
where $i$ in $\left(R_\nu^\mu\right)_j^i$ represents the number of Ricci tensors, and $j$ represents the number of possible ways to contract the $i$ number of Ricci tensors. Generically, each higher curvature combination has a different dimensionful coupling constant denoted by $a_{n i}^j$. This theory generically has many maximally symmetric vacua, many massive degrees of freedom, and is nonunitary at the tree level about its constant curvature vacua. However, due to the determinantal nature of the Lagrangian, BINMG (\ref{BINMG}) as a very specific example of the generic theory (\ref{Gen}) is consistent at infinite order or at any finite truncation in powers of curvature. By consistency, we mean it reproduces, up to desired order in the curvature expansion, the extended NMG theories that are consistent with the AdS/CFT duality and that have a c-function \cite{Tekin2, Sinha, Paulos}. See \cite{Sisman} for details of these calculations. 
\item The BINMG action appears as a counterterm in AdS4 \cite{Sinha2} and might have a supersymmetric extension \cite{Bergshoeff}.
\end{enumerate}
All these considerations encourage us to study the Euclidean version of the BINMG theory following \cite{Visser1}.

Going back to the partition function, for the spatial volume constraint (here it is actually an area of the disk, yet we will use the word volume to conform with the higher-dimensional cases) we have $C(g) = \int d^2 x \sqrt{\gamma}$ and $\mu = V$ where the induced metric on the spatial disk is 
\begin{equation}
\gamma_{\mu\nu} := g_{\mu\nu}-N^2 \left(\partial_{\mu}\phi\right) \left(\partial_{\nu}\phi\right), \hskip 1.5 cm N := \left(g^{\mu\nu}\left(\partial_{\mu}\phi\right)\left(\partial_{\nu}\phi\right)\right)^{-\frac{1}{2}}.
\end{equation}
Therefore, in the end, the constraint can be written as
\begin{equation}
\frac{1}{\hbar} \int d\phi \xi(\phi) \left(\int d^2 x \sqrt{\gamma}-V\right),
\end{equation}
which reduces the partition function to
\begin{eqnarray}
\mathcal{Z}(V,\Lambda,m^2) = \int\int \mathcal{D} g \mathcal{D} \xi  e^{-I_{E}(g)+\frac{1}{\hbar} \int d\phi \xi (\phi) \left(\int d^2 x \sqrt{\gamma}-V\right)}.
\end{eqnarray}
In the next section, we will evaluate this partition function in the saddle point approximation, which is dominated by constrained instantons. For this purpose, we first need to find the field equations.
\subsection{Field equations}
When the constraint equation is satisfied, $C(g)=V$, the field equations can be obtained from the variation of the action at the saddle point approximation,
\begin{equation}
\delta I_{E}(g) - \frac{1}{\hbar}\int d\phi \xi(\phi) \delta (C(g)-\mu) = 0.
\end{equation}
From the constraint equation, we get
\begin{equation}
\delta\left(\frac{1}{\hbar} \int d \phi \xi(\phi)(C(g)-\mu)\right)= \frac{1}{\hbar} \int d \phi \xi(\phi) \delta C(g)
\end{equation}
and
\begin{eqnarray}
\delta C(g) &=& \int d^2x \delta (\sqrt{\gamma}) = \int d^2x \left(-\frac{1}{2N} \sqrt{g} \gamma_{\mu\nu} \delta g^{\mu\nu}\right).
\end{eqnarray}
The field equations of BINMG are rather cumbersome, but fortunately, they were found in \cite{Tekin2}. Following that computation verbatim, one first writes the Euclidean form of the BINMG action (\ref{BINMG}) in a more convenient way as
\begin{equation}
I_E=-\frac{4 m^2}{\kappa^2} \int d^3x \sqrt{\det(g)} F(R,K,S),
\end{equation}
where the Lagrangian is
\begin{eqnarray}
F(R,K,S) &:=& \sqrt{1-\frac{\sigma}{2 m^2}\left(R+\frac{\sigma}{m^2}K-\frac{1}{12 m^4}S\right)}-\left(1-\frac{\lambda_0}{2}\right),
\end{eqnarray}
and the curvature invariants are defined as
\begin{equation}
K := R_{\mu\nu} R^{\mu\nu}-\frac{1}{2} R^{2}, \hspace{1.5 cm} S := 8 R^{\mu\nu}R_{\mu\alpha}R^{\alpha}_{\nu}-6 R R_{\mu\nu} R^{\mu\nu}+R^{2}.  
\end{equation}
Hence, one can get the variation of the action as
\begin{eqnarray}
\delta I_E(g)&=& -\frac{4m^2}{\kappa^2}\int d^3x \left(\delta\sqrt{\det(g)} F + \sqrt{\det(g)} \delta F\right)\nonumber\\
&=&-\frac{4m^2}{\kappa^2}\int d^3x \sqrt{\det(g)} \left(-\frac{1}{2}g_{\mu\nu}F\delta g^{\mu\nu}+ \frac{\partial F}{\partial R}\delta R+\frac{\partial F}{\partial K}\delta K +\frac{\partial F}{\partial S} \delta S\right).
\end{eqnarray}
The final form of the action variation including the constraint part is
\begin{eqnarray}
-\frac{4m^2}{\kappa^2}\int d^3x \sqrt{\det(g)} \delta g^{\mu\nu} \left(-\frac{1}{2} g_{\mu\nu} F+\frac{\partial F}{\partial R}\frac{\delta R}{\delta g^{\mu\nu}}+\frac{\partial F}{\partial K}\frac{\delta K}{\delta g^{\mu\nu}}+\frac{\partial F}{\partial S}\frac{\delta S}{\delta g^{\mu\nu}}-\frac{\kappa^2}{8 m^2 \hbar}\frac{\tilde{\lambda}(\phi)}{N}\gamma_{\mu\nu}\right),
\end{eqnarray}
which yields the field equations
\begin{equation}
\mathcal{E}_{\mu\nu}=\frac{\kappa^2}{8 m^2 \hbar} T_{\mu\nu},
\end{equation}
where
\begin{equation}
\mathcal{E}_{\mu\nu} := -\frac{1}{2} g_{\mu\nu} F+\frac{\partial F}{\partial R}\frac{\delta R}{\delta g^{\mu\nu}}+\frac{\partial F}{\partial K}\frac{\delta K}{\delta g^{\mu\nu}}+\frac{\partial F}{\partial S}\frac{\delta S}{\delta g^{\mu\nu}},
\end{equation}
and the constraint acts as a source of perfect fluid without an energy density but with pressure
\begin{equation}
T_{\mu\nu} := \frac{\xi(\phi)}{N} \gamma_{\mu\nu}.
\end{equation}
To be able to find the constrained instanton solution, we still need the explicit form of the field equations which were given in \cite{Tekin2} as  
\begin{eqnarray}
&&\frac{\kappa^2}{8 m^2 \hbar^2} T_{\mu \nu}= \nonumber \\
 && -\frac{1}{2} F g_{\mu \nu}+\left(g_{\mu \nu} \square-\nabla_\mu \nabla_\nu\right) F_R+F_R R_{\mu \nu}-\frac{\sigma}{m^2}\left\{2 \nabla_\alpha \nabla_\mu\left(F_R R^\alpha{ }_\nu\right)-g_{\mu \nu} \nabla_\beta \nabla_\alpha\left(F_R R^{\alpha \beta}\right)\right. \nonumber\\
&& \left.-\square\left(F_R R_{\mu \nu}\right)-2 F_R R_\nu{ }^\alpha R_{\mu \alpha}+g_{\mu \nu} \square\left(F_R R\right)-\nabla_\mu \nabla_\nu\left(F_R R\right)+F_R R R_{\mu \nu}\right\}-\frac{1}{2 m^4}\left\{4 F_R R^\rho{ }_\mu R_{\rho \alpha} R^\alpha{ }_\nu\right.\nonumber \\
&& +2 g_{\mu \nu} \nabla_\alpha \nabla_\beta\left(F_R R^{\beta \rho} R^\alpha{ }_\rho\right)+2 \square\left(F_R R_\nu{ }^\rho R_{\mu \rho}\right)-4 \nabla_\alpha \nabla_\mu\left(F_R R_\nu{ }^\rho R^\alpha{ }_\rho\right)+2 \nabla_\alpha \nabla_\mu\left(F_R R R^\alpha{ }_\nu\right) \nonumber\\
&& -g_{\mu \nu} \nabla_\alpha \nabla_\beta\left(F_R R R^{\alpha \beta}\right)-\square\left(F_R R R_{\mu \nu}\right)-2 F_R R R_\nu{ }^\rho R_{\mu \rho}-g_{\mu \nu} \square\left(F_R R_{\alpha \beta}^2\right)+\nabla_\nu \nabla_\mu\left(F_R R_{\alpha \beta}^2\right)\nonumber \\
&& \left.-F_R R_{\alpha \beta}^2 R_{\mu \nu}+\frac{1}{2} g_{\mu \nu} \square\left(F_R R^2\right)-\frac{1}{2} \nabla_\mu \nabla_\nu\left(F_R R^2\right)+\frac{1}{2} F_R R^2 R_{\mu \nu}\right\}, \label{field}
\end{eqnarray}

where 
\begin{equation}
F_R := \frac{\partial F}{\partial R} = - \frac{\sigma}{4 m^2 \left[F+\left(1-\frac{\lambda}{2}\right)\right]}.
\end{equation}
Next, we will solve these equations for a particular saddle point that has two Killing symmetries. From now on, we will work in the units for which $\hbar=c=1$.
\subsection{Saddle point metric}
Let us consider a metric with $\phi$ and $\theta$ as Killing coordinates such that $\tilde{\lambda}(\phi)=\tilde{\lambda}$:
\begin{equation}
ds^2 := N^2(r) d\phi^2 + h(r) dr^2 + r^2 d\theta^2. \label{metric}
\end{equation}
$N(r)$ and $h(r)$ are to be determined from (\ref{field}) and the boundary conditions to be discussed. When (\ref{metric}) is plugged into (\ref{field}), the resulting differential equations are still highly complicated. Therefore, guided by the discussion in \cite{Visser1}, we take the following ansatz:
\begin{eqnarray}
h(r) &:=& \frac{1}{1-\frac{r^2}{L^2}}=\frac{1}{1-\Lambda r^2},
\end{eqnarray}
where $L$ is the dS radius defined as $L^2=1/\Lambda$.
Given this $h(r)$ in this saddle point metric, one can determine $N(r)$, but the ordinary differential equation is still highly nonlinear, and it pays to discuss the boundary conditions first.  We expect the lapse function $N(r)$ to vanish on the surface of the constraint volume $N(R_V)=0$, where $R_V$ can be found by using the constraint equation $\int d^2x \sqrt{\gamma}=V$. Analogous to the discussion given in \cite{Visser1}, to remove the canonical singularity at the surface of the constraint boundary, we impose 
\begin{equation}
\left.\frac{dN}{dl} \right\vert_{l=0}=1,
\end{equation}
where $l$ is the distance from the horizon, i.e., the constraint volume surface $l=R_V-r$.
\section{$\lambda_0=0$ Solution}
For the flat spacetime, $\lambda_0=0$, $\Lambda = 0$ and $h(r)=1$. When we insert (\ref{metric}) into (\ref{field}), one finds that in order to obtain $\mathcal{E}_{\phi\phi}=0$ since $T_{\phi\phi}=0$, the lapse function should be of the form $N(r) = \alpha r^2 + \beta $, where $\alpha$ and $\beta$ are to be found from the boundary conditions. The first boundary condition, namely, the lapse function should vanish on the surface of the volume constraint, gives $N(r) = \alpha \left(r^2-R_V^2\right)$. For the second boundary condition, we have 
\begin{equation}
\left.\frac{dN(l)}{dl}\right\vert_{l=0}=\left(2\alpha l -2\alpha R_V \right)_{l=0}=-2\alpha R_V=1
\end{equation}
yielding $N(r) = \frac{1}{2 R_V}\left(R_V^2-r^2\right)$. From the constraint equation, we have
\begin{equation}
\int d^2x \sqrt{\det(\gamma)}=\int_{0}^{R_V}\int_{0}^{2\pi}dr d\theta \, r=\pi R_V^2=V,
\end{equation}
so $R_V = \sqrt{\frac{V}{\pi}}$. In conclusion, the metric we obtained for the flat spacetime at the Euclidean saddle becomes
\begin{eqnarray}
ds^2 &=& \frac{1}{4 R_V^2}\left(R_V^2-r^2\right)^2 d\phi^2 + dr^2 + r^2 d\theta^2.
\end{eqnarray}
Finally, for this constrained instanton, we get
\begin{equation}
\sqrt{\det(g+\frac{\sigma}{m^2}G)} = \frac{r\left(r^2-R_V^2+\frac{2}{m^2}\right)}{2R_V}, \hskip 1.2 cm \sqrt{\det(g)} = \frac{r\left(r-R_V\right)\left(r+R\right)}{2R_V},
\end{equation}
and the action becomes
\begin{eqnarray}
I_E=- \frac{4m^2}{\kappa^2}\int d^3x \left[\sqrt{\det(g+\frac{\sigma}{m^2}G)}-\sqrt{\det(g)}\right]= -\frac{8 \pi^2}{\kappa^2} R_V.
\end{eqnarray} 
The circumference of the disk has the ``area'' $A_V=2\pi R_V$.
As a result, the action in terms of the area can be written as
\begin{equation}
I_E = - \frac{8 \pi^2}{\kappa^2} R_V = \left(-\frac{4\pi}{\kappa^2}\right) A_V = - \frac{A_V}{4 G_3}, \label{entropy1}
\end{equation}
where in the last equality, we introduced the three-dimensional Newton's constant $G_3$ as
\begin{equation}
G_3 = \frac{\kappa^2}{16 \pi} .
\end{equation}
This result is the same as the Bekenstein-Hawking or Gibbons-Hawking results, except here we calculated not for a black hole or de Sitter horizon but for a bounded disk.  Hence, the dimension of the Hilbert space is $\mathcal{Z} = \exp\left(\frac{A_V}{4 G_3}\right)$.

As a final step, for completeness, let us calculate the Lagrange multiplier $\xi$. We know that 
\begin{equation}
\mathcal{E}_{rr}=\frac{1}{m^2\left(R_V^2-r^2\right)}, \hskip 1.2 cm T_{rr}=\frac{2 R_V}{R_V^2-r^2} \xi.
\end{equation}
This component of the field equations yields
\begin{eqnarray}
\xi &=& \frac{1}{4 \pi G_3 R_V}.
\end{eqnarray}
\section{$\lambda_0 \ne 0$ Solution}
We again take the metric as (\ref{metric}) with $h(r)=\frac{1}{1-\frac{r^2}{L^2}}$.
Let us make a change of variables to simplify the ensuing discussion and define $r=L \sin\chi$ and $u=\sin\chi$ for which the metric becomes
\begin{equation}
ds^2 = L^2 N^2 \left(u\right) d\phi^2 + \frac{L^2}{1-u^2} du^2 + L^2 u^2 d\theta^2.
\end{equation}
We assume $N(\chi) = \alpha \cos\chi + \beta$. The first boundary condition on the lapse function discussed in the previous section gives $N(\chi) = \alpha \left(\cos\chi-\cos\chi_V\right)$. From the second boundary condition, one determines $\alpha=\frac{1}{\sin\chi_V}$.
As a result, the lapse function becomes
\begin{equation}
N\left(\chi\right) = \frac{\cos\chi-\cos\chi_V}{\sin\chi_V}, \hskip 1.2 cm \text{so} \hskip 1.2 cm N\left(u\right)=\frac{\sqrt{1-u^2}-\sqrt{1-u_V^2}}{u_V},
\end{equation}
and the metric becomes
\begin{eqnarray}
ds^2 =&& L^2 \left(\frac{\sqrt{1-u^2}-\sqrt{1-u_V^2}}{u_V}\right)^2 d\phi^2 + \frac{L^2}{1-u^2} du^2 + L^2 u^2 d\theta^2. \label{metric2}
\end{eqnarray}
We still need to determine the relation between $L^2$, $m^2$, and the bare cosmological constant $\lambda_0$. This follows from $\mathcal{E}_{\phi\phi} = \frac{\kappa^2}{8 m^2 \hbar} T_{\phi\phi}=0$, and we get
\begin{equation}
\lambda_0 = -\frac{2 L m}{\sqrt{L^2 m^2-\sigma }}+\frac{2 \sigma }{L m \sqrt{L^2 m^2-\sigma }}+2,
\end{equation}
or in a more compact form,
\begin{equation}
\frac{1}{L^2}=\Lambda = m^2 \sigma \lambda_0 \left(1-\frac{\lambda_0}{4}\right),
\end{equation}
which is the same result obtained in \cite{Tekin2}.  The next order of business is to evaluate the action of the constrained instanton metric (\ref{metric2}). In the $u$ coordinate the action looks a little cumbersome, but in the $\chi$ coordinate, it becomes
\begin{eqnarray}
I_E = -\frac{4 m^2}{\kappa^2} \int d^3x \left[-\frac{\sigma  \sin \chi  \sqrt{L^2 m^2-\sigma }}{m^3 \sin\chi_V}\right] &=& \frac{4 \sigma \sqrt{L^2 m^2-\sigma}}{\kappa^2 m \sin\chi_V} \int_{0}^{2\pi} \int_{0}^{\chi_V} \int_{0}^{2 \pi} d\phi d\chi d\theta \sin\chi\nonumber\\
&=&\frac{\pi \sigma \sqrt{L^2 m^2-\sigma}}{m G_3 \sin\chi_V} \left(1-\cos\chi_V\right), \label{actionv}
\end{eqnarray}
where in the last equality we used $\kappa^2 = 16 \pi G_3$.
Now, we can turn to our constraint equation and calculate the ``volume'' and area of the disk. For the volume of the boundary, one has
\begin{eqnarray}
V&=&\int d^2x \sqrt{\gamma}=2 \pi L^2 \left(1-\cos\chi_V\right),
\end{eqnarray}
and the area of the boundary becomes
\begin{eqnarray}
A_V&=& \int_{0}^{2\pi} d\theta L \sin\chi_V = 2 \pi L \sin\chi_V.
\end{eqnarray}
In terms of these, the action (\ref{actionv}) becomes
\begin{eqnarray}
I_E&=&\left(\frac{\pi}{G_3}\frac{V}{A_V}\right)\sigma\sqrt{1-\frac{\sigma}{L^2m^2}}. \label{actionv2}
\end{eqnarray}
For a generic $n$-ball, one has $V=\frac{R_V}{2}A_V=\frac{L\sin\chi_V}{2}A_V$. Therefore, (\ref{actionv2}) becomes
\begin{eqnarray}
I_E &=&\frac{A_V}{4G_3}\sigma\sqrt{1-\frac{\sigma \Lambda}{m^2}}. 
\end{eqnarray}
This is an interesting result: For $\Lambda=0$ and $\sigma=-1$, it reduces to (\ref{entropy1}). Therefore, from now on we set $\sigma=-1$. For $\Lambda \ne 0$, this is exactly the same as the Wald entropy \cite{Wald} defined as 
\begin{equation}
S_W = - 2 \pi \oint \left(\frac{\partial L}{\partial R_{\rho\sigma}}g^{\mu\nu}\varepsilon_{\mu\rho}\varepsilon_{\nu\sigma}\right)d^3x,
\end{equation}
where $\varepsilon_{\mu\rho}$ is the binormal vector to the constraint boundary surface. It is shown in \cite{Tekin3} that the Wald entropy for the BINMG in the de Sitter spacetime becomes \footnote{Note that in \cite{Tekin3} the overall sign of the Wald entropy is opposite because the action in that work is written with the pseudo-Riemannian metric.}
\begin{equation}
S_W=\frac{A}{4}\frac{1}{G_3}\sqrt{1+\frac{\Lambda}{m^2}}.
\end{equation}
Therefore, the Euclidean action of the constrained instanton is exactly equal to the minus of the Wald entropy evaluated not at the de Sitter horizon but at the constraint boundary. Moreover, we can also relate the result to the Bekenstein-Hawking or Gibbons-Hawking result as follows. In the BINMG theory, one needs to define the effective gravitational constant as \cite{Tekin3, Brustein}
\begin{equation}
\frac{1}{G_{\text{eff}}} :=\frac{1}{G_3}\frac{\bar{R}_{\mu\nu}}{\bar{R}}\left(\frac{\partial \mathcal{L}}{\partial R_{\mu\nu}}\right)_{\bar{R}}=\frac{1}{G_3}\sqrt{1+\frac{\Lambda}{m^2}}.
\end{equation}
Hence, as a result, the Wald entropy is just like the Bekenstein-Hawking entropy
\begin{equation}
S_W=\frac{A}{4 G_{\text{eff}}}.
\end{equation}
In conclusion, the entropy that we have calculated for the constrained volume is equivalent to the Wald or Bekenstein-Hawking entropies, and the dimension of the Hilbert space is given as
\begin{equation}
\mathcal{Z}\left(\Lambda, V, m^2\right) = e^{-I_E}= e^{\frac{A_V}{4G_3}\sqrt{1+\frac{\Lambda}{m^2}}}=e^{\frac{A_V}{4 G_{\text{eff}}}}.
\end{equation}
Let us note that, even though the constrained instanton has a finite action, as in the case of \cite{Visser1}, for both $\lambda_0=0$ and $\lambda_0 \ne 0$ cases, there is a mild singularity at the boundary for the Ricci tensor and the stress-energy tensor.

As a final step, for completeness, let us calculate the Lagrange multiplier $\xi$. By using the relation
\begin{equation}
\mathcal{E}_{\mu\nu} = \frac{\kappa^2}{8 m^2} T_{\mu\nu} = \frac{2 \pi G_3}{m^2} \frac{\xi}{N(\chi)} \gamma_{\mu\nu}, 
\end{equation}
and the explicit forms of $\mathcal{E}_{rr}$ and $\gamma_{rr}$, one can find 
\begin{eqnarray}
\xi &=& -\frac{\cot\chi_V}{4 \pi G_3 \sqrt{L^2 m^2+1}} = -\frac{1}{4 \pi G_3} \frac{1}{m R_V}\frac{\cos\chi_V}{\sqrt{1+\frac{\Lambda}{m^2}}}.
\end{eqnarray}
\section{Conclusions and Discussion}
Following Jacobson and Visser \cite {Visser1}, who generalized the notion of the gravitational entropy of horizons to the constrained finite spatial volume of $n$-ball topologies in (cosmological) Einstein gravity, we computed the quantum gravity partition function at the saddle point approximation in an infinite-order gravity theory of the  Born-Infeld form in $2+1$ dimensions. The theory has 2 propagating degrees of freedom at the perturbative level and is locally nontrivial, unlike three-dimensional Einstein gravity. Our computations support the conclusions of \cite{Visser1}: The dimension of the Hilbert space of all states of a given disk with a boundary is equal to the exponential of the Bekenstein-Hawking, Gibbons-Hawking, or Wald entropy of it. These results are encouraging: In a coming work, we shall give a similar analysis in the $n$-dimensional Born-Infeld gravity theory that has the same perturbative and vacuum structure as Einstein's theory \cite{Tekin4, Tekin5, Tekin6}.

\end{document}